\documentclass[twoside,twocolumn,english,pre]{revtex4}
\usepackage[T1]{fontenc}
\usepackage[latin1]{inputenc}
\usepackage{float}
\usepackage{amsmath}
\usepackage{graphicx}

\makeatletter
\usepackage{babel}
\makeatother
\begin{document}

\preprint{revised PRE Rapid Communication publication version}

\title{Generalized Berry Conjecture and mode correlations in chaotic plates}

\author{Alexei Akolzin}

\author{Richard L. Weaver}

\email{r-weaver@uiuc.edu}

\affiliation{Department of Theoretical and Applied Mechanics, University of Illinois,
104 S. Wright Street, Urbana, Illinois 61801, USA}

\begin{abstract}
We consider a modification of the Berry Conjecture for eigenmode statistics
in wave-bearing systems. The eigenmode correlator is conjectured to
be proportional to the imaginary part of the Green's function. The
generalization is applicable not only to scalar waves in the interior
of homogeneous isotropic systems where the correlator is a Bessel
function, but to arbitrary points of heterogeneous systems as well.
In view of recent experimental measurements, expressions for the intensity
correlator in chaotic plates are derived.
\end{abstract}

\date{\today}

\maketitle
In 1977 Berry conjectured that the higher eigenmodes of a ray-chaotic
Hamiltonian, in particular, a billiard, should be statistically indistinguishable
from a superposition of plane standing waves of all directions, with
uncorrelated amplitudes and phases \cite{cite:Berry.77}; the idea
is also found in \cite{cite:Ebeling.et.al}. One immediate consequence
is that the modes are Gaussian random functions, with correlations
given by \[
\left\langle u^{\left(n\right)}\left(\mathbf{x}\right)u^{\left(m\right)}\left(\mathbf{x}+\mathbf{r}\right)\right\rangle =A^{2}\delta_{nm}J_{0}\left(kr\right),\]
 where brackets $\left\langle \cdot\right\rangle $ represent spatial
average over position $\mathbf{x}$, \textbf{}$A$ \textbf{}is an
unimportant normalization factor, $r=\left|\mathbf{r}\right|$ is
a separation distance, and \textbf{$k$} is the wavenumber that appears
in the governing Helmholtz equation, $\left(\nabla^{2}+k^{2}\right)u^{\left(n\right)}\left(\mathbf{x}\right)=0$.
Another immediate consequence is that the intensity correlator is\begin{equation}
\left\langle u^{\left(n\right)}\left(\mathbf{x}\right)^{2}u^{\left(n\right)}\left(\mathbf{x}+\mathbf{r}\right)^{2}\right\rangle =A^{4}\left[1+2J_{0}^{2}\left(kr\right)\right].\label{eq:ScalarIntensityCorrelator}\end{equation}

Berry established the conjecture for an asymptotic regime in which
wavelengths are much less than system size. It is only approximate
at finite wavelength, i.e, in practice. The conjecture has been shown
to be incorrect at finite wavelength, inasmuch as many modes show
evidence of scarring \cite{cite:Heller.84}, existence of which may
be understood in the light of the {}``quantum equidistribution theorem''
put forth by Shnirelman \cite{cite:Shnirelman}. The conjecture is
manifestly incorrect if attention is restricted to points near a boundary
where, locally, plane waves are correlated with their reflections.
Nevertheless, numerical and experimental evidence shows that it is
widely satisfied \cite{cite:experimental}, and references in \cite{cite:Stockmann}. 

Recent measurements on elastic waves in plates have underlined the
inadequacy of the conjecture, as stated, for systems more complicated
than the scalar billiard \cite{cite:Schaadt.03}. Three dimensional
microwave billiards (or merely thick quasi 2D billiards) for which
the electric field satisfies a vector wave equation, and elastic wave
systems in general, require a statement about the correlations of
the vector-valued eigenmodes \cite{cite:Eckhardt.99}. Even in scalar
wave systems, if they are inhomogeneous, or if interest includes points
near boundaries, the conjecture needs modification. It is not sufficient
to extend the conjecture by expressing the modes as uncorrelated superpositions
of plane waves of all wave types, as their relative amplitudes remain
unspecified. In those recent measurements on elastic waves in plates,
the intensity correlator did not satisfy (\ref{eq:ScalarIntensityCorrelator}).
This note is intended to provide the appropriate generalization of
Berry's conjecture needed for experiments in wave-bearing systems
more complex than scalar billiards.

We begin with an identity, written in the form it takes for a tensor
Green's function appropriate for a vector wave equation, \begin{equation}
\mathbf{G}\left(\mathbf{x}_{1},\mathbf{x}_{2},\omega\right)=\sum_{n}\frac{\mathbf{u}^{\left(n\right)}\left(\mathbf{x}_{1}\right)\otimes\mathbf{u}^{\left(n\right)}\left(\mathbf{x}_{2}\right)}{\omega_{n}^{2}-\left(\omega+\imath\varepsilon\right)^{2}},\label{eq:GreenDecomposition}\end{equation}
 as a modal sum over the normalized real modes with eigenfrequencies
$\omega_{n}$. The imaginary part of this Green's dyadic is \cite{cite:Economou}\[
\Im\mathbf{G}\left(\mathbf{x}_{1},\mathbf{x}_{2},\omega\right)=\frac{\pi}{2\omega}\sum_{n}\mathbf{u}^{\left(n\right)}\left(\mathbf{x}_{1}\right)\otimes\mathbf{u}^{\left(n\right)}\left(\mathbf{x}_{2}\right)\delta\left(\omega-\omega_{n}\right).\]
 This may be averaged, either over a short range in frequency, or
over an ensemble of systems that differ from the system of interest
only at positions far from the closely spaced points $\mathbf{x}_{1}$
and $\mathbf{x}_{2}$. In either case $\mathbf{G}$ is largely unaffected.
The right side becomes the corresponding modal correlator. It is seen
to be in general not $J_{0}$, but rather $\Im\mathbf{G}$. It is
not a spatial average, as called for by the Berry conjecture, but
rather a frequency or ensemble average.

Thus we are led to a generalized Berry conjecture. Based upon the
exact identity for ensemble or frequency averages, we conjecture that
it is also true for spatial averages at a fixed mode. This reduces
to the Berry conjecture for the simple case of a scalar wave. The
conjecture about the correlator is, as demonstrated above, manifestly
correct if what one means by the averaging is a frequency or ensemble
average. The potentially more problematic aspects lie in the supposition
that this correlator may be found within spatial averages on a single
mode of a single sample from the ensemble, or for that matter that
the statistics are Gaussian.

When this generalized Berry conjecture is applied to the modes of
an infinite isotropic homogeneous 3D elastic body, the modal correlator
is given by the Green's function, $\mathbf{G}^{\infty}$, satisfying
Navier's equations \cite{cite:Graff}: \[
\left(\lambda+\mu\right)\nabla\left(\nabla\cdot\mathbf{G}^{\infty}\right)+\mu\nabla^{2}\mathbf{G}^{\infty}+\rho\omega^{2}\mathbf{G}^{\infty}=-\mathbf{I}\delta^{3}\left(\mathbf{r}\right),\]
 with $\mathbf{I}$ being the identity tensor, $\lambda$ and $\mu$
being elastic Lam\'{e} constants, and $\rho$ representing material
density. The spatial Fourier transform of the solution is obtained
readily:\[
\rho\mathbf{G}^{\infty}\left(\mathbf{q},\omega\right)=\frac{\mathbf{q}\otimes\mathbf{q}/\left|\mathbf{q}\right|^{2}}{\left|\mathbf{q}\right|^{2}c_{l}^{2}-\left(\omega+\imath\varepsilon\right)^{2}}+\frac{\mathbf{I}-\mathbf{q}\otimes\mathbf{q}/\left|\mathbf{q}\right|^{2}}{\left|\mathbf{q}\right|^{2}c_{t}^{2}-\left(\omega+\imath\varepsilon\right)^{2}}.\]
 The first term is the longitudinal part with wavespeed $c_{l}$,
the second is the transverse part with wavespeed $c_{t}$. On taking
the imaginary part of the 3D inverse Fourier transform, one finds
\begin{align*}
\Im\mathbf{G}^{\infty}\propto\int & \cos\left(\mathbf{q}\cdot\mathbf{r}\right)\Bigl[c_{l}^{-3}\delta\left(\left|\mathbf{q}\right|-\omega/c_{l}\right)\mathbf{q}\otimes\mathbf{q}/\left|\mathbf{q}\right|^{2}\\
+ & c_{t}^{-3}\delta\left(\left|\mathbf{q}\right|-\omega/c_{t}\right)\left(\mathbf{I}-\mathbf{q}\otimes\mathbf{q}/\left|\mathbf{q}\right|^{2}\right)\Bigr]d^{2}\Omega_{\mathbf{q}}dq.\end{align*}
It is seen that $\Im\mathbf{G}^{\infty}$ is a superposition of plane
waves of the two types and of all directions of propagation, with
relative strengths given by the inverse cubes of the wave speeds,
i.e. by equipartition \cite{cite:Weaver.82}.

If the frequency averaging is over a sufficiently broad band, then,
even in a finite system, the correlator reduces to that in the unbounded
medium, $\Im\mathbf{G}^{\infty}$. This is readily established, as
in \cite{cite:Weaver.86}, by recognizing that short time responses
are independent of distant parts of a structure, and that sufficiently
short time responses are equivalent to frequency averaging over bands
of sufficient width. The theorem is readily generalized to the vicinity
of a boundary, or a scatterer. In these cases it is not difficult
to show that the diffuse field-field correlator may be also constructed
by superposing an equipartitioned set of uncorrelated incident plane
or incoming or standing waves together with their coherent reflections
and scatterings. For the special case of elastic waves near a free
surface, this was explored in \cite{cite:Weaver.85.et.al}.

The Green's function is more complicated in a plate, in particular
if its thickness is comparable to a wavelength. In the work reported
by Schaadt \emph{et al.} \cite{cite:Schaadt.03}, an intensity correlator
was constructed by averages over space and over a small number of
modes. Due to the good preservation of up/down symmetry, all modes
were either of a purely flexural (odd up/down parity) character, or
a mixture of extensional and shear (even up/down parity). Thus their
correlator should be the imaginary parts of the partial Green's functions.
Except for the effects of nonzero thickness to wavelength ratio, these
modes have displacements that are purely out-of-plane or purely in-plane
respectively.

Modes which are antisymmetric on reflection about the mid-plane, sometimes
called flexural, are un-coupled to the others; at frequencies below
the first cutoff ($\omega=\pi c_{t}/2h$, where $h$ is the half-width
of the plate) there is a single wave number $k_{f}$ governing such
waves. Their intensity correlator must therefore be formed from a
single Bessel function $J_{0}\left(k_{f}r\right)$. This was in fact
observed \cite{cite:Schaadt.03}; Schaadt \emph{et al.} reported a
good fit to $1+2J_{0}^{2}\left(k_{f}r\right)$.   These modes in fact
have vector valued fields, so the correlator is technically a tensor.
 The measured correlator is a contraction of that tensor with the
(unknown) polarization vector $\hat{\mathbf{p}}$ of the detector.
Given a field (an eigenmode) $\psi\left(\mathbf{x}\right)$ they constructed
an 'intensity correlator' \[
I\left(r\right)=\overline{\left\langle \left[\hat{\mathbf{p}}\cdot\psi\left(\mathbf{x}\right)\right]^{2}\left[\hat{\mathbf{p}}\cdot\psi\left(\mathbf{x}+\mathbf{r}\right)\right]^{2}\right\rangle \Bigm/\left\langle \left[\hat{\mathbf{p}}\cdot\psi\left(\mathbf{x}\right)\right]^{2}\right\rangle ^{2}},\]
 where the overbar indicates additional average over the direction
of the vector $\mathbf{r}$. We presume that the detector polarization
$\hat{\mathbf{p}}$ is held fixed during this averaging. Inasmuch
as $\mathbf{u}$ is a Gaussian process, the above fourth order statistic
reduces to the sum of three products of two second order statistics.
In terms of the generalized Berry conjecture the correlator is rewritten
as \begin{equation}
I\left(\mathbf{r}\right)=1+2\overline{\left\langle \hat{\mathbf{p}}\cdot\Im\mathbf{G}\left(\mathbf{x},\mathbf{x}+\mathbf{r}\right)\cdot\hat{\mathbf{p}}\right\rangle ^{2}}\bigm/\left\langle \hat{\mathbf{p}}\cdot\Im\mathbf{G}\left(\mathbf{x},\mathbf{x}\right)\cdot\hat{\mathbf{p}}\right\rangle ^{2}.\label{eq:IntensityCorrelator}\end{equation}
 For the surface of a plate in flexure, even far from the edges, such
that $\mathbf{G}\approx\mathbf{G}^{\infty}$, the Green's function
is not as simple as might have been hoped, \begin{alignat*}{1}
\Im\mathbf{G}^{\infty}\left(\mathbf{r}\right)\propto & \hat{\mathbf{x}}_{1}\otimes\hat{\mathbf{x}}_{1}\nu^{2}\left[J_{0}\left(k_{f}r\right)-J_{2}\left(k_{f}r\right)\right]/2\\
+ & \hat{\mathbf{x}}_{2}\otimes\hat{\mathbf{x}}_{2}\nu^{2}\left[J_{0}\left(k_{f}r\right)+J_{2}\left(k_{f}r\right)\right]/2\\
+ & \left(\hat{\mathbf{x}}_{1}\otimes\hat{\mathbf{x}}_{3}-\hat{\mathbf{x}}_{3}\otimes\hat{\mathbf{x}}_{1}\right)\nu J_{1}\left(k_{f}r\right)\\
+ & \hat{\mathbf{x}}_{3}\otimes\hat{\mathbf{x}}_{3}J_{0}\left(k_{f}r\right).\end{alignat*}
 Unit vectors $\hat{\mathbf{x}}_{1}$ and $\hat{\mathbf{x}}_{2}$
lie in the mid-plane of the plate, with $\hat{\mathbf{x}}_{1}$ taken
in the direction of $\mathbf{r}$, and $\hat{\mathbf{x}}_{3}$ is
normal to the plane and pointing towards the surface at hand. Factor
$\nu$ represents a degree of in-plane surface motion associated with
such waves, and vanishes at long wavelength. The correlator $I$ is
not only dependent upon separation distance $r$, but also upon the
angle between polarization of the detector $\hat{\mathbf{p}}$ and
the separation vector direction $\mathbf{r}/r$.

\begin{figure}[H]
\includegraphics{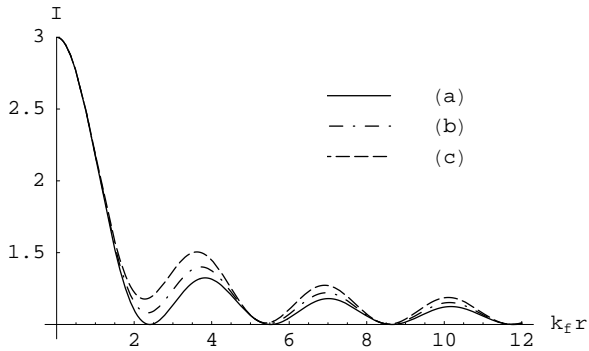}

\caption{\label{cap:flexCorrelator} Intensity correlator of a single flexural
mode for $\nu\rho$ equal (a) $0$, (b) $2$, and (c) $\infty$.}
\end{figure}

By averaging over the direction of vector $\mathbf{r}$, for a single
flexural mode, one finds (see Appendix) \begin{equation}
I\left(r\right)=1+2\frac{J_{0}^{2}\left(k_{f}r\right)\left(1+\nu^{2}\rho^{2}/2\right)^{2}+J_{2}^{2}\left(k_{f}r\right)\nu^{4}\rho^{4}/8}{\left(1+\nu^{2}\rho^{2}/2\right)^{2}},\label{eq:IntensityCorrelatorSingle}\end{equation}
 with $\rho^{2}=\left(p_{x}^{2}+p_{y}^{2}\right)/p_{z}^{2}$. The
correlator is plotted in Figure \ref{cap:flexCorrelator} for a number
of values of $\nu\rho$. %
 It is equal to $3$ at zero separation, $r=0$, as demanded by Gaussian
mode statistics, and is higher than $1+2J_{0}^{2}\left(k_{f}r\right)$
for non-zero values of $\nu\rho$, with the most pronounced difference
observed near the first minimum.

At realistic values of $\nu\rho$ (Schaadt \emph{et al.} estimate
$\rho\sim0.33$; we calculate $\nu=-0.68$ at the relevant frequencies),
the effect is small, and difficult to resolve within the data's precision.
The sole anomaly in the data is the best-fit value of the relative
variance at zero separation: $2.93\pm0.05$, an anomaly with only
small statistical significance. Such a value cannot be explained with
the current theory; indeed the basic assumption of Gaussian statistics
demands that this quantity be $3$. However, if the quantity $2.93$
is understood as the ratio between the relative variance at zero separation
and at the first minimum, then the current theory can explain the
anomaly, by calling for $\left|\nu\rho\right|=1.06$.

The data's precision does not support any more detailed comparisons.
This is also the case with the in-plane modes. Modes which have even
up/down parity consist of an equipartitioned diffuse mixture of longitudinal
waves with wavenumber $k_{l}$ (which have both in-plane and, due
to Poisson effect, out-of-plane components of displacement), and in-plane
horizontally polarized shear waves with wavenumber $k_{sh}$.  These
waves mode convert to one and other at the plate boundaries.  Thus
the relevant Green's tensor has two wavenumbers, and one anticipates
structures like those seen in Figs. 4 and 5 of Schaadt \emph{et al}.
 However, as in the flexural case, one does not expect to see simple
Bessel functions $J_{0}$, but rather also terms in $J_{1}$ and $J_{2}$.
The relative amplitudes of these several terms are not obvious \emph{a
priori} but could be predicted by the present theory. An attempt to
fit their data to the present theory is probably unwarranted at this
time. A revisit to structures like theirs, but with a well characterized
detector of known polarization, may be indicated.

In summary, we have advanced a modification of Berry Conjecture, appropriate
for the eigenmode statistics of wave-bearing systems. It is expected
to be relevant, not only for elastic waves in homogeneous plates,
but in general statistical physics of waves in heterogeneous and mode-converting
systems as well.

Acknowledgment: this work was supported by the National Science Foundation
CMS-0201346.

\appendix

\section{Multi-mode intensity correlator in a chaotic plate}

We start calculation of the full intensity correlator by first considering
the normal modes of the Rayleigh-Lamb spectrum \cite{cite:Graff}.
The displacement vector of these modes is given by\[
\mathbf{u}=\left[U\left(x_{3}\right)\left(k^{-1}\nabla\right)+\hat{\mathbf{x}}_{3}W\left(x_{3}\right)\right]f\left(x_{1},x_{2}\right),\]
 with $f$ satisfying a scalar 2D Helmholtz equation: $\left[\nabla^{2}+k^{2}\right]f\left(x_{1},x_{2}\right)=0$.
The displacement components $U$ and $W$ are the solutions of a boundary-value
ODE in $x_{3}$. With the vertical wavenumbers of longitudinal and
shear waves defined as $\alpha^{2}=\omega^{2}/c_{l}^{2}-k^{2}$, and
$\beta^{2}=\omega^{2}/c_{t}^{2}-k^{2}$ , one deduces the dispersion
relation for the odd and even up/down parity modes\[
\tan\beta h\bigm/\tan\alpha h=-\left[\left(k^{2}-\beta^{2}\right)^{2}\bigm/4\alpha\beta k^{2}\right]^{\pm1},\]
 where $+1$ in the exponent corresponds to the odd parity modes,
and $-1$ to the even modes. The dispersion relation gives the wavenumbers
of the odd ($k_{f}$) and even modes ($k_{l}$) as multi-branched
implicit functions of the frequency: $k=k_{n}\left(\omega\right)$.
Expressions for $U$ and $W$ of the odd and even modes respectively
are 

\begin{align*}
U= & 2k^{3}\beta\sin\beta h\sin\alpha x_{3}-\left(k^{2}-\beta^{2}\right)k\beta\sin\alpha h\sin\beta x_{3},\\
W= & 2k^{2}\alpha\beta\sin\beta h\cos\alpha x_{3}+\left(k^{2}-\beta^{2}\right)k^{2}\sin\alpha h\cos\beta x_{3};\end{align*}

\begin{align*}
U= & 2k^{3}\beta\cos\beta h\cos\alpha x_{3}-\left(k^{2}-\beta^{2}\right)k\beta\cos\alpha h\cos\beta x_{3},\\
W= & 2k^{2}\alpha\beta\cos\beta h\sin\alpha x_{3}+\left(k^{2}-\beta^{2}\right)k^{2}\cos\alpha h\sin\beta x_{3}.\end{align*}

By specifying a complete set of solutions $f$ in the plane (for example,
standing plane waves or standing cylindrical waves), we construct
the modes of an infinite plate. Alternatively, we may specify a complete
set of propagating waves $f$, in which case a complex conjugate must
be inserted on the first factor $\mathbf{u}$ in equation (\ref{eq:GreenDecomposition}).
These modes are not the natural modes of a finite plate unless the
boundary conditions at the outer rim are particularly special. They
may nevertheless be used in a modal expansion of the Green's function
if attention is confined to early enough times (alternatively if a
frequency averaging is done) as discussed above. The average of the
exact Green's function, $\mathbf{G}$, can then be substituted by
the Green's function in the infinite plate, $\mathbf{G}^{\infty}$. 

We construct a partial Green's function (\ref{eq:GreenDecomposition})
of the Rayleigh-Lamb spectrum, and find its imaginary part, \begin{align}
\Im G_{\alpha\beta}^{\infty}= & \sum_{n}\biggl[a_{n}J_{0}\left(k_{n}r\right)\delta_{\alpha\beta}/2\nonumber \\
 & \quad+b_{n}J_{2}\left(k_{n}r\right)\left(\delta_{\alpha\beta}/2-r_{\alpha}r_{\beta}/r^{2}\right)\biggr],\nonumber \\
\Im G_{33}^{\infty}= & \sum_{n}c_{n}J_{0}\left(k_{n}r\right),\label{eq:GreenComponents}\\
\Im G_{\alpha3}^{\infty}= & -\Im G_{3\alpha}=\sum_{n}d_{n}J_{1}\left(k_{n}r\right)r_{\alpha}/r.\nonumber \end{align}
 The sum is taken over propagating, i.e. having real $k_{n}$, modes
only. Greek indices span the in-plane space: $\alpha,\beta=\left\{ 1,2\right\} $.
By means of the factor, $\nu_{n}=U\left(h\right)/W\left(h\right)|_{k=k_{n}\left(\omega\right)}$,
we can write the modal amplitudes, $a_{n}=b_{n}=c_{n}\nu_{n}^{2}$,
and $d_{n}=c_{n}\nu_{n}$, in terms of the amplitude describing out-of-plane
displacement of the plate surface, \[
c_{n}=\frac{\pi}{4}\frac{\partial k}{\partial\omega}\frac{k}{\omega}\frac{W^{2}\left(h\right)}{\int_{-h}^{+h}\left[U^{2}\left(x_{3}\right)+W^{2}\left(x_{3}\right)\right]dx_{3}}\biggr|_{k=k_{n}\left(\omega\right)}.\]

The horizontal shear modes have displacements purely in the plane
of the plate: \[
\mathbf{u}=V\left(x_{3}\right)\left(k^{-1}\nabla\right)\times\left[\hat{\mathbf{x}}_{3}f\left(x_{1},x_{2}\right)\right],\]
 the dispersion relation for the shear wavenumbers ($k_{sh}$) being:
$k=k_{n}\left(\omega\right)=\sqrt{\left(\omega/c_{t}\right)^{2}-\left(\pi n/2h\right)^{2}}$.
Imaginary part of the corresponding partial Green's function has the
same form as for the Rayleigh-Lamb modes (\ref{eq:GreenComponents}).
However, the modal amplitudes are now as follows, $a_{n}=-b_{n}=\left(1+\delta_{0n}\right)/4hc_{t}^{2}$,
and $c_{n}=d_{n}=0$.

The full multi-mode tensor Green's function includes the modes of
all (namely, odd and even parity Rayleigh-Lamb, and horizontal shear)
branches required for its short-time expansion at a given frequency.
The propagating modes of these branches contribute to the full intensity
correlator (\ref{eq:IntensityCorrelator}), \begin{widetext}\[
I\left(r\right)=1+2\frac{\left[\sum_{n}\left(a_{n}\rho^{2}/2+c_{n}\right)J_{0}\left(k_{n}r\right)\right]^{2}+\left[\sum_{n}b_{n}J_{2}\left(k_{n}r\right)\right]^{2}\rho^{4}/8}{\left[\sum_{n}\left(a_{n}\rho^{2}/2+c_{n}\right)\right]^{2}}.\]
 \end{widetext} The averages over directions of the separation vector
$\mathbf{r}$ are carried out with the help of the following rules:\begin{align*}
\left\langle r_{\alpha}r_{\beta}/r^{2}\right\rangle \textrm{=} & \delta_{\alpha\beta}/2,\\
\left\langle r_{\alpha}r_{\beta}r_{\gamma}r_{\iota}/r^{4}\right\rangle = & \left(\delta_{\alpha\beta}\delta_{\gamma\iota}+\delta_{\alpha\gamma}\delta_{\beta\iota}+\delta_{\alpha\iota}\delta_{\beta\gamma}\right)/8\end{align*}

In the special case that we have, the frequency is such that there
is only one odd (flexural) mode, and the sum is replaced with a single
term, yielding correlator (\ref{eq:IntensityCorrelatorSingle}). The
factor $\nu$ is computed for the plate parameters of Ref. \cite{cite:Schaadt.03}
(thickness $3\,\textrm{mm}$, Poisson ratio $0.33$, transverse wavespeed,
$c_{t}=3.1\,\textrm{mm}/\mu\textrm{s}$), and excitation frequencies
$432$, $510$, $513$, $514\,\textrm{kHz}$, to be $\nu=-0.68$.
\end{document}